\documentclass[unnumsec,webpdf,modern,large]{mam-authoring-template}%

\usepackage{amsmath,amssymb,bm}
\usepackage{etoolbox}
\usepackage{placeins}

\graphicspath{{figure/}}

\theoremstyle{thmstyleone}%
\theoremstyle{thmstyletwo}%
\theoremstyle{thmstylethree}%

\newlength{\EqVSpace}
\newcommand{\ApplyEquationSpacing}{%
    \setlength{\abovedisplayskip}{\EqVSpace}%
    \setlength{\belowdisplayskip}{\EqVSpace}%
    \setlength{\abovedisplayshortskip}{\EqVSpace}%
    \setlength{\belowdisplayshortskip}{\EqVSpace}%
}

\newcommand{\ii}{\mathrm{i}}
\newcommand{\rmpi}{\mathrm{\pi}}
\newcommand{\rvec}{\mathbf{r}}
\newcommand{\kvec}{\mathbf{k}}
\newcommand{\Rp}{\mathbf{R}_{\mathrm{p}}}
\newcommand{\Qp}{\mathbf{Q}_{\mathrm{p}}}
\newcommand{\FT}{\mathcal{F}}

\setcitestyle{authoryear,round}

\newlength{\symmetricmargin}
\makeatletter
\patchcmd{\@maketitle}
  {{{\sffamilyfontcnitalic\fontsize{8bp}{12}\selectfont \@journaltitle}, \sffamilyfontcn\@copyrightyear, \thepage--\thelastpage}}
  {\mbox{}}
  {}
  {\PackageWarning{3DLINV}{Could not suppress the first-page journal/page-range line}}
\patchcmd{\@maketitle}
  {{\sffamilyfontcn doi: \@DOI}}
  {\mbox{}}
  {}
  {\PackageWarning{3DLINV}{Could not suppress the first-page DOI line}}
\patchcmd{\@maketitle}
  {{\color{black!20}\rule{45pt}{55pt}}}
  {}
  {}
  {}

\def\addressandsep{}
\def\addresscommasep{\par}

\def\ps@opening{%
  \let\@mkboth\@gobbletwo
  \def\@oddhead{}%
  \def\@evenhead{}%
  \def\@oddfoot{\hfil\thepage\hfil}%
  \let\@evenfoot\@oddfoot
}
\makeatother

\begin{document}

\journaltitle{}
\DOI{}
\copyrightyear{}

\title[3D reconstruction of weakly scattering objects]{Three-dimensional reconstruction of weakly scattering objects from single-defocus 4D-STEM data}

\author[1,2$\ast$]{Kousuke Ooe\ORCID{0000-0001-9632-3821}}
\author[3]{Yufan Shen\ORCID{0009-0001-2097-6911}}
\author[3]{Yuichi Shimakawa\ORCID{0000-0003-1019-2512}}
\author[3,4]{Daisuke Kan\ORCID{0000-0002-7505-0059}}
\author[1]{Scott D. Findlay\ORCID{0000-0003-4862-4827}}

\authormark{Ooe et al.}

\address[1]{\orgdiv{School of Physics and Astronomy}, \orgname{Monash University}, \orgaddress{\street{Clayton}, \state{Victoria} \postcode{3800}, \country{Australia}}}
\address[2]{\orgdiv{Nanostructures Research Laboratory}, \orgname{Japan Fine Ceramics Center}, \orgaddress{\street{Nagoya}, \state{Aichi} \postcode{456-8587}, \country{Japan}}}
\address[3]{\orgdiv{Institute for Chemical Research}, \orgname{Kyoto University}, \orgaddress{\street{Uji}, \state{Kyoto} \postcode{611-0011}, \country{Japan}}}
\address[4]{\orgdiv{Graduate School of Engineering}, \orgname{The University of Osaka}, \orgaddress{\street{Suita}, \state{Osaka} \postcode{565-0871}, \country{Japan}}}

\corresp[$\ast$]{Corresponding author. \href{mailto:kousuke.ooe@monash.edu}{kousuke.ooe@monash.edu}}

\abstract{
Three-dimensional (3D) atomic-structure analysis is of central importance in (scanning) transmission electron microscopy (STEM). STEM techniques to reconstruct 3D structure include tilt-series or defocus-series annular dark-field, best suited to heavier atomic species, and dynamical-scattering-model inversion applied to single-defocus four-dimensional (4D) STEM data, requiring computationally expensive post-experiment analysis. In this study, we propose a 3D-structure linear inversion (3D LINV) STEM algorithm, which enables a fast and direct reconstruction of 3D structures from a single-defocus 4D-STEM dataset. This method is based on solving linear equations relating the measured 4D-STEM data and the unknown 3D specimen potential under the thick weak phase object approximation. We show the capabilities of 3D LINV STEM through systematic 4D-STEM simulations across various observation parameters such as specimen thickness, partial coherence, and shot noise, and through an experimental demonstration of 3D structure reconstruction of a stacked freestanding membrane with atomic-scale lateral resolution and nanometre-scale depth resolution.
}

\keywords{4D-STEM, 3D structure reconstruction, linear scattering theory, contrast transfer function}

\maketitle
\ApplyEquationSpacing

\section{Introduction}\label{introduction}

Three-dimensional (3D) atomic structure characterisation is a central interest of (scanning) transmission electron microscopy ((S)TEM), particularly for specimens with inhomogeneous structures along the depth direction, such as crystals with point defects and dopants, heterostructured thin-films, and stacked two-dimensional materials. While the lateral resolution of STEM has improved thanks to the development of aberration correctors \citep{haider1998,batson2002}, which is powerful for analysing two-dimensional (2D), i.e. depth-projected, structures, 3D structure characterisation requires resolving the structure along depth direction. Perhaps the two main established approaches for 3D characterisation in (S)TEM are tilt-series tomography and focus-series depth sectioning, in which multiple (S)TEM images over a range of specimen tilts (tomography) or defocus values (depth sectioning) are sequentially acquired to reconstruct the 3D structure \citep{borisevich2006,miao2016}. In terms of atomic-resolution STEM, the tomographic approach has been mainly used for visualising 3D structure of nanoparticles \citep{zhou2019} whereas the depth-sectioning technique has mainly been used for stacked membranes \citep{kim2025} or bulk specimens with dopants \citep{ishikawa2020}. Typically, these techniques rely on high-angle annular dark-field (HAADF) STEM imaging, which offers straightforwardly interpretable image contrast at heavier atomic sites due to its incoherent imaging characteristics.

More recently, four-dimensional (4D) STEM \citep{ophus2019} has emerged as a versatile platform for materials characterisation, not only in 2D but also in 3D. 4D STEM collects 2D diffraction patterns at every probe position during a 2D raster scan of the electron probe, forming a 4D dataset as a function of probe position in real space and detector coordinate in reciprocal space. 4D STEM enables synthetic virtual detector imaging \citep{kimoto2011,pennycook2015,esser2023}, electromagnetic-field imaging \citep{muller2014,krajnak2016}, crystal orientation mapping \citep{jeong2021,ophus2022}, strain mapping \citep{ozdol2015,pekin2017}, and phase reconstruction \citep{rodenburg2008,miao2025}. In particular, phase reconstruction techniques such as electron ptychography have been intensively utilised for atomic-scale structure retrievals. In the literature, ptychographic reconstructions initially assumed projection approximations, including the phase object approximation (POA) for Wigner distribution deconvolution (WDD) ptychography \citep{yang2017} and extended ptychographic iterative engine (ePIE) \citep{maiden2009}, and the weak phase object approximation (WPOA) for single side band (SSB) ptychography \citep{pennycook2015}. As an extension of the WPOA, the thick WPOA (tWPOA) \citep{seki2018} has been used for optimum bright-field (OBF) reconstruction \citep{ooe2021}. While these methods can visualise the atomic sites of both heavy and light elements due to phase contrast scaling approximately linearly with atomic number, they mainly capture depth-averaged structural information\footnote{As a notable exception to this, \citep{yang2016} performed depth sectioning by setting multiple defocus values for the deconvolution kernel in the WDD ptychographic reconstruction process based on single-defocus 4D-STEM data.}. For resolving a depth-dependent structure by 4D-STEM, structure reconstruction methods may employ dynamical scattering models such as the multislice model \citep{gao2017,chen2021} and the scattering-matrix model \citep{brown2022}. While these methods can reconstruct 3D atomic structures from single/multiple-defocus 4D dataset(s) experimentally \citep{oleary2024,schloz2025,kp2025}, structure inversion processes incorporating dynamical scattering models tend to be computationally expensive, often taking a few hours to obtain 3D reconstructions \citep{chen2021,lee2025}. Furthermore, these methods may need finely tuned hyperparameters such as learning rates and amplitudes of regularisations to get well converged results after iterative calculations \citep{lee2025}.

In this study, we propose a 3D linear-inversion (3D LINV) STEM algorithm, enabling 3D structure reconstruction from a single-defocus 4D-STEM dataset via inversion of the tWPOA model. The 3D LINV STEM algorithm can rapidly and directly reconstruct 3D structures from a single-defocus 4D-STEM dataset without computationally expensive iterative calculations, and with minimal hyperparameters or regularisation, making it powerful to analyse structures nonuniform along the depth direction. This paper is structured as follows. We first revisit the tWPOA model by introducing a five-dimensional (5D) contrast transfer function (CTF) and establish a matrix-based linear equation to be solved for the unknown 3D structure. We investigate the capabilities of this 3D LINV method through 3D structure reconstruction from simulated 4D-STEM datasets, clarifying how accurately the linear model can reconstruct 3D structure in the presence of multiple scattering. In addition, the reconstruction results are investigated under a series of thicknesses, convergence angles, defocuses, and partial coherence. We then introduce a noise normalisation scheme to the reconstruction from finite-dose 4D-STEM data with shot noise, which improves visualising the reconstructed atomic structure. Finally, we apply our 3D LINV approach to an experimental 4D-STEM dataset acquired from a stacked membrane specimen, demonstrating the capability to reconstruct the 3D structure in a practical environment.

\section{Theoretical framework}\label{sec:theory}

\subsection{The forward scattering model}\label{subsec:forward}

The STEM wavefield in vacuum may be written
\begin{equation}
\psi_{\mathrm{in}}(\rvec,\Rp;z)
=
\FT^{-1}_{\kvec\rightarrow\rvec}
\left[
T(\kvec;z)\exp\!\left(-2\rmpi\ii\,\kvec\cdot\Rp\right)
\right],
\label{eq:psi-in}
\end{equation}
where \(\mathbf{r}\) is the two-dimensional real-space coordinate, \(\mathbf{k}\) is its reciprocal space conjugate, z is taken to be zero (\(z = 0\)) at the focal plane and positive along the electron-incident direction, \(\Rp\) is the probe position, \(\mathcal{F}_{\mathbf{k} \rightarrow \mathbf{r}}^{- 1}\) denotes inverse Fourier transform, and \(T\left( \mathbf{k;}z \right) = A\left( \mathbf{k} \right)\exp\left\lbrack - \ii\chi\left( \mathbf{k};z \right) \right\rbrack\) is a lens transfer function, where \(A\left( \mathbf{k} \right)\) is a top-hat function equal to 1 for \(\left| \mathbf{k} \right| \leq k_{0}\) (with \(k_{0}\) the reciprocal-space aperture cutoff) and 0 elsewhere, and \(\chi\left( \mathbf{k};z \right) = \pi\lambda\left| \mathbf{k} \right|^{2}z + \chi^{'}\left( \mathbf{k} \right)\) is the aberration function with the first term indicating defocus (\(z\) denoting a propagation distance from the focal plane) and the second including the other aberrations.

\begin{figure*}[h!]
\centering
\includegraphics[width=0.905\textwidth]{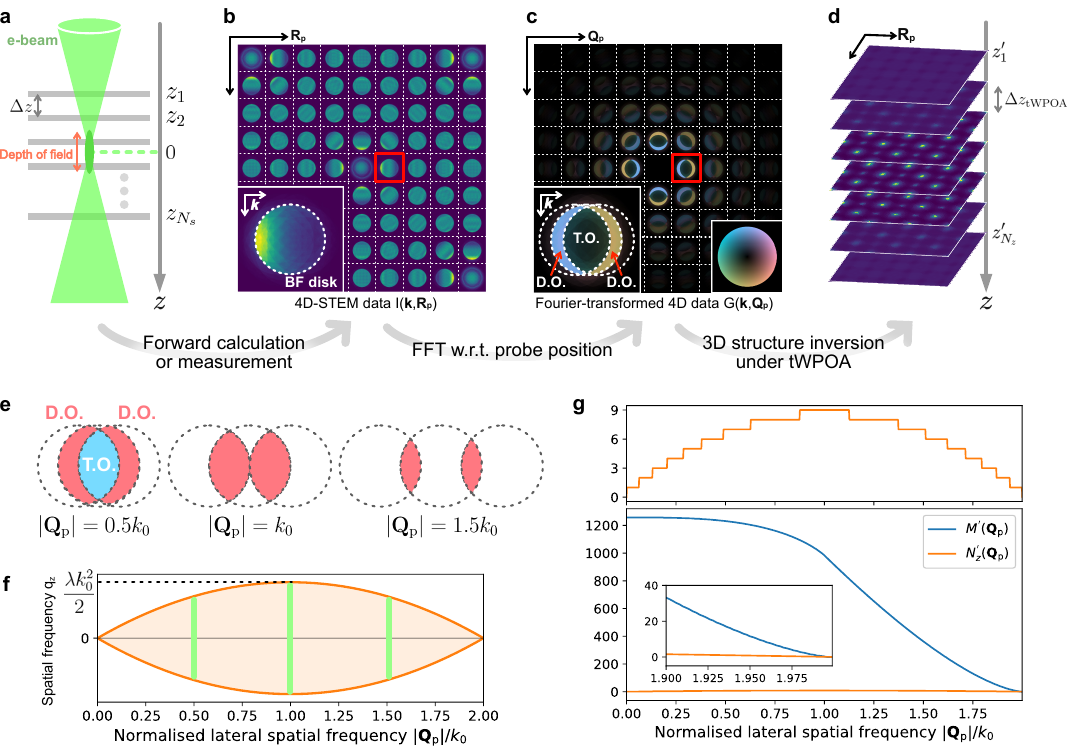}
\caption{Schematic of the workflow in 3D LINV STEM. (a) Schematic of the multislice model, resulting in (b) the 4D-STEM intensity, from which we calculate (c) its Fourier transform with respect to the probe position (a complex 4D distribution for which amplitude and phase are shown as per the inset colour wheel). White-dotted lines separate different probe position in (b) and spatial frequency coordinates in (c). The insets magnify the tile highlighted by the red box. (d) Reconstructed 3D phase distribution. (e) Schematic illustration of double-overlap (D.O.) and triple-overlap (T.O.) regions between a transmitted disk and singly-scattered disks at three different \(\left| \Qp \right|\) values. (f) Region of non-zero 5D CTF (orange shading), with solid lines showing the information limit along \(q_{z}\) as a function of \(\left| \Qp \right|\). Vertical green lines indicate the range at different \(\left| \Qp \right|\) values. (g) The number of sampling points along \(\left\{ \mathbf{k}^{\mathbf{'}} \right\}\) and \(\left\{ q_{z}^{'} \right\}\) coordinates (\(M^{'}\left( \Qp \right)\) and \(N_{z}^{'}\left( \Qp \right)\), respectively) as a function of \(\left| \Qp \right|\). The upper plot shows \(N_{z}^{'}\left( \Qp \right)\) alone and the lower plot shows both together. The inset magnifies the plots around the lateral information limit about \(\left| \Qp \right|\sim 2k_{0}\). These plots assume a square area spanning the bright-field disk sampled by 40\(\times\)40 pixels and an accelerating voltage and a convergence semi-angle of 300 kV and 40 mrad, respectively. The specimen-thickness range for 3D reconstruction is assumed to be 10 nm.\emph{\hfill\break
}}
\label{fig:1}
\end{figure*}

Under the multislice model shown schematically in Fig.~\ref{fig:1}a, a full dynamical scattering model in STEM, the exit-surface wavefunction \(\psi_{\mathrm{exit}}\left( \rvec,\Rp\mathbf{;}z \right)\) may be written
\begin{equation}
\begin{aligned}
\psi_{\mathrm{exit}}(\rvec,\Rp;z_1+t)
={}&O_{N_s}(\rvec)\widehat{P}_{\Delta z}\cdots
\Bigl[
O_2(\rvec)\widehat{P}_{\Delta z}
\\[-1pt]
&\qquad\times
\Bigl[O_1(\rvec)\psi_{\mathrm{in}}(\rvec,\Rp;z_1)\Bigr]
\Bigr],
\end{aligned}
\label{eq:multislice}
\end{equation}
where \(z_{1}\) is the depth of the sample surface, \(t = \left( N_{s} - 1 \right)\mathrm{\Delta}z\) is the sample thickness (for \(N_{s}\) thin slices of thickness \(\mathrm{\Delta}z\) as shown in Fig.~\ref{fig:1}a), \(O_{j}\left( \mathbf{r} \right) = \exp\left( \ii\sigma V_{j}\left( \mathbf{r} \right) \right)\ \left( j = 1,2,\ldots,N_{s} \right)\) is a transmission function with \(\sigma\) an interaction constant and \(V_{j}\left( \mathbf{r} \right)\) the projected potential of the \(j\)-th slice, and \({\widehat{P}}_{\Delta z}\lbrack \bullet \rbrack = \mathcal{F}_{\mathbf{k} \rightarrow \mathbf{r}}^{- 1}\left\lbrack \exp\left( - \ii\rmpi\lambda\left| \mathbf{k} \right|^{2}\Delta z \right)\mathcal{F}_{\mathbf{r} \rightarrow \mathbf{k}}\lbrack \bullet \rbrack \right\rbrack\) denotes a propagator with propagation distance \(\Delta z\) between slices. Having defined the origin of the z coordinate as the focal plane, the \emph{z} coordinate of the sample surface also equals the conventional defocus value \(\Delta f = z_{1}\), with the convention \(\Delta f > 0\) for overfocus and \(\Delta f < 0\) for underfocus.

If the phase shift due to the electrostatic potential is sufficiently small, such that \(O_{j}\left( \mathbf{r} \right) \cong 1 + \ii\sigma V_{j}\left( \mathbf{r} \right)\), the exit wavefunction \(\psi_{\mathrm{exit}}\left( \rvec,\Rp\mathbf{;}z_{1} + t \right)\) can be approximated as
\begin{equation}
\begin{aligned}
\psi_{\mathrm{exit}}(\rvec,\Rp;z_1+t)
={}&
\widehat{P}_{(N_s-1)\Delta z}
\Bigl[\psi_{\mathrm{in}}(\rvec,\Rp;z_1)\Bigr]
\\
&+\ii\sigma\sum_{j=1}^{N_s}
\widehat{P}_{(N_s-j)\Delta z}
\Bigl[
V_j(\rvec)\widehat{P}_{(j-1)\Delta z}
\\[-1pt]
&\hspace{8.0em}\times
\Bigl[\psi_{\mathrm{in}}(\rvec,\Rp;z_1)\Bigr]
\Bigr],
\end{aligned}
\label{eq:twpoa-wave}
\end{equation}
where the first term corresponds to unscattered electrons and the second term to the superposition of electrons singly-scattered at each slice \(j\). Neglecting multiple scattering (defined here to be any terms beyond linear order in the scattering potential) but accounting for all propagation inside the specimen is called the thick weak phase object approximation (tWPOA) \citep{seki2018,ooe2019,seki2022}.

The intensity of 4D-STEM data \(I\left( \kvec,\Rp \right)\), as a function of reciprocal space coordinates and probe position, is given by
\begin{equation}
I(\kvec,\Rp)
=
\left|
\FT_{\rvec\rightarrow\kvec}
\left[
\psi_{\mathrm{exit}}(\rvec,\Rp;z_1+t)
\right]
\right|^2.
\label{eq:intensity}
\end{equation}
For notational simplicity, we omit the explicit dependence of the intensity on \(z_{1}\) and \(t\) unless it is necessary for clarity. A simulated example dataset is shown in Fig.~\ref{fig:1}b. Assuming the tWPOA, substituting Eq.~\eqref{eq:twpoa-wave} into Eq.~\eqref{eq:intensity} and Fourier transforming with respect to probe position gives
\begin{equation}
\begin{aligned}
G(\kvec,\Qp)
&=
\FT_{\Rp\rightarrow\Qp}\!\left[I(\kvec,\Rp)\right]
\\
&=
|A(\kvec)|^2\delta(\Qp)
+
\sum_{j=1}^{N_s}
\beta(\kvec,\Qp;z_j)\Phi_j(\Qp),
\end{aligned}
\label{eq:g-function}
\end{equation}
where \(\Qp\) is the spatial frequency conjugate to \(\Rp\), \(\Phi_{j}\left( \Qp \right) = \FT_{\rvec\rightarrow\Qp}\left\lbrack \phi_{j}\left( \mathbf{r} \right) \right\rbrack\) is the Fourier transform of the phase shift due to the \(j\)-th potential slice \(\phi_{j}\left( \mathbf{r} \right)\) (i.e., \(\phi_{j}\left( \mathbf{r} \right) = \sigma V_{j}\left( \mathbf{r} \right)\) with \(\sigma\) denoting an interaction constant), and
\begin{equation}
\begin{aligned}
\beta(\kvec,\Qp;z_j)
=\ii\Bigl[&
T^{*}(\kvec;z_j)T(\kvec-\Qp;z_j)
\\[-1pt]
&-T(\kvec;z_j)T^{*}(\kvec+\Qp;z_j)
\Bigr]
\end{aligned}
\label{eq:beta-4dctf}
\end{equation}
is the four-dimensional contrast transfer function (4D-CTF) for slice \(z_{j}\) \citep{ooe2025,varnavides2026}, which describes how the information of the specimen \(\Phi_{j}\left( \Qp \right)\) is transferred into \(G\left( \kvec,\Qp \right)\) at each spatial frequency \(\Qp\) and detector coordinate \(\mathbf{k}\). The function \(G\left( \kvec,\Qp \right)\) is a four-dimensional complex quantity. Fig.~\ref{fig:1}c shows \(G\left( \kvec,\Qp \right)\) calculated from the simulated 4D dataset \(I\left( \kvec,\Rp \right)\) in Fig.~\ref{fig:1}b. For simplicity, we assume that aberrations other than defocus are corrected (i.e., \(\chi^{'}\left( \mathbf{k} \right) = 0\) in Eq.~\eqref{eq:psi-in}) in this study, reducing Eq.~\eqref{eq:beta-4dctf} to
\begin{equation}
\begin{aligned}
\beta(\kvec,\Qp;z_j)
=\ii A(\kvec)\Bigl[&
A(\kvec-\Qp)e^{-2\rmpi\ii\omega_-z_j}
\\[-1pt]
&-A(\kvec+\Qp)e^{-2\rmpi\ii\omega_+z_j}
\Bigr],
\end{aligned}
\label{eq:beta-defocus}
\end{equation}
where
\begin{equation}
\omega_{\pm}
=
-\lambda\left(
\kvec\cdot\Qp
\pm
\frac{|\Qp|^2}{2}
\right).
\label{eq:omega}
\end{equation}
The terms \(\omega_{\pm}z_{j}\) denote the degree of interference between the direct-beam disk and scattered disks shifted by \(\pm \Qp\) in the presence of the defocus \(z_{j}\) \citep{nogami2026}.

To clarify the contrast transfer nature along the z direction, we express \(\Phi_{j}\left( \Qp \right)\) as a Fourier series
\begin{equation}
\Phi_j(\Qp)
=
\sum_{l=1}^{N_s}
\widehat{\Phi}_{q_{z,l}}(\Qp)
\exp\!\left(2\rmpi\ii q_{z,l}z_j\right),
\label{eq:phi-fourier}
\end{equation}
where \(\left\{ q_{z,l} \right\}_{l = 1}^{N_{s}}\) is the frequency coordinate conjugate to the real-space z-coordinate \(\left\{ z_{j} \right\}_{j = 1}^{N_{s}}\). Substituting Eq.~\eqref{eq:phi-fourier} into Eq.~\eqref{eq:g-function} (and henceforth ignoring the \(\Qp=0\) term since \(\beta\left( \mathbf{k,}\Qp\mathbf{= 0;}z_{j} \right) = 0\) means \(\Phi_{j}\left( \Qp=0 \right)\) cannot be reconstructed) gives
\begin{align}
G(\kvec,\Qp)
&=
\sum_{j=1}^{N_s}
\beta(\kvec,\Qp;z_j)
\sum_{l=1}^{N_s}
\widehat{\Phi}_{q_{z,l}}(\Qp)
\exp\!\left(2\rmpi\ii q_{z,l}z_j\right)
\nonumber\\
&=
\sum_{l=1}^{N_s}
\widehat{\beta}(\kvec,\Qp,q_{z,l})
\widehat{\Phi}_{q_{z,l}}(\Qp),
\label{eq:g-5dctf}
\end{align}
where
\begin{equation}
\widehat{\beta}(\kvec,\Qp,q_{z,l})
=
\sum_{j=1}^{N_s}
\beta(\kvec,\Qp;z_j)
\exp\!\left(2\rmpi\ii q_{z,l}z_j\right).
\label{eq:beta-hat-def}
\end{equation}
As the modified 4D-CTF \(\widehat{\beta}\left( \kvec,\Qp,q_{z,l} \right)\) in Eq.~\eqref{eq:beta-hat-def} works as a transfer function from the Fourier coefficients of the phase \({\widehat{\Phi}}_{q_{z,l}}\left( \Qp \right)\) to \(G\left( \kvec,\Qp \right)\) as a function of \(\mathbf{k}\), \(\Qp\) and \(q_{z,l}\) (compare the forms of Eq.~\eqref{eq:g-5dctf} and Eq.~\eqref{eq:g-function}), we call it a 5D-CTF in this study. Substituting Eq.~\eqref{eq:beta-defocus} into Eq.~\eqref{eq:beta-hat-def} gives
\begin{equation}
\begin{aligned}
\widehat{\beta}(\kvec,\Qp,q_{z,l})
={}&\ii A(\kvec)\Bigl[
A(\kvec-\Qp)W(q_{z,l}-\omega_-)
\\[-1pt]
&\hspace{4.2em}-A(\kvec+\Qp)W(q_{z,l}-\omega_+)
\Bigr],
\end{aligned}
\label{eq:5dctf}
\end{equation}
where
\begin{equation}
W(q_{z,l}-\omega_{\pm})
=
\sum_{j=1}^{N_s}
e^{2\rmpi\ii(q_{z,l}-\omega_{\pm})z_j}.
\label{eq:W}
\end{equation}
If the sampling of \(\mathbf{k}\), \(\Qp\) and hence \(\omega_{\pm}\) are commensurate with that of \(\left\{ q_{z,l} \right\}\), Eq.~\eqref{eq:W} reduces to a (Kronecker) delta function that returns a constant at \(q_{z,l} = \omega_{\pm}\) and zero elsewhere. Otherwise, this function becomes a so-called Dirichlet kernel \citep{press2007}, which is still highly peaked about \(q_{z,l} = \omega_{\pm}\) but has a finite-width distribution, as described in more detail in Supplementary Note 2. Together with Eq.~\eqref{eq:5dctf} and for a given \(\Qp\) value, this puts an effective upper bound on the magnitude of \(q_{z}\) for non-zero 5D-CTF across all \(\mathbf{k}\) values of
\begin{equation}
q_{z,\max}(\Qp)
=
\lambda\left(
 k_0|\Qp|-
 \frac{|\Qp|^2}{2}
\right),
\qquad
0\le |\Qp|\le 2k_0.
\label{eq:qz-limit}
\end{equation}
This constitutes an information limit on the 5D-CTF along the \(q_{z}\) axis, indicating that \(\widehat{\beta}\left( \kvec,\Qp,q_{z,l} \right)\) with \(q_{z,l}\) components outside this limit (i.e., \(\left| q_{z,l} \right| > q_{z,\max}\left( \Qp \right)\) at each \(\Qp\)) is zero and the corresponding Fourier coefficients of the phase \({\widehat{\Phi}}_{q_{z,l}}\left( \Qp \right)\) do not contribute to \(G\left( \kvec,\Qp \right)\), i.e. do not affect the measured 4D STEM intensity (within the tWPOA). We note that information limit in Eq.~\eqref{eq:qz-limit} is consistent with the bounds on the 3D contrast transfer function for depth sectioning via WDD ptychography \citep{yang2016} and for depth sectioning via ADF STEM \citep{xin2009}, meaning that larger convergence angles and shorter electron wavelengths (i.e., higher accelerating voltages) enable better depth resolution via a higher information limit along \(q_{z}\).

\subsection{The inverse problem}\label{subsec:inverse}

Under the tWPOA model, Eq.~\eqref{eq:g-5dctf} shows that \(G\left( \kvec,\Qp \right)\), obtainable by Fourier transform over measured 4D-STEM data at a single defocus, is linearly related to the Fourier components of the 3D scattering potential \({\widehat{\Phi}}_{q_{z,l}}\left( \Qp \right)\) (in phase units) with the coefficients being the 5D-CTF \(\widehat{\beta}\left( \kvec,\Qp,q_{z,l} \right)\), which is calculable for characterised lens properties and detector sampling. Here, we seek to solve these equations for \({\widehat{\Phi}}_{q_{z,l}}\left( \Qp \right)\) to reconstruct the unknown three-dimensional phase distribution from measured 4D-STEM data. The form of Eq.~\eqref{eq:g-5dctf} means different values of \(\Qp\) are completely independent of one another, but that the value of \(G\left( \kvec,\Qp \right)\) for fixed \(\Qp\) depends on the values of \({\widehat{\Phi}}_{q_{z,l}}\left( \Qp \right)\) for all spatial frequency \(q_{z,l}\). For a given value of \(\Qp\), the possibility and stability of solving the corresponding equations depends on whether the number of \(\mathbf{k}\) values at which \(G\left( \kvec,\Qp \right)\) is non-zero exceeds the number of \(q_{z,l}\) values at which we seek to determine \({\widehat{\Phi}}_{q_{z,l}}\left( \Qp \right)\).

Assume the diffraction pattern is discretely sampled on \(M\) pixels with coordinates \(\left\{ \mathbf{k} \right\} = \mathbf{k}_{1},\ldots\mathbf{k}_{M}\). It is known that the 4D-CTF \(\beta\left( \kvec,\Qp;z_{j} \right)\ \)may have non-zero values inside the double and triple overlap regions between the direct-beam disk \(A\left( \mathbf{k} \right)\) and two scattered disks shifted by \(\pm \Qp\) but will be zero elsewhere \citep{pennycook2015,yang2015,ooe2025,nogami2026}, as depicted in Fig.~\ref{fig:1}e. The same applies to the 5D-CTF \(\widehat{\beta}\left( \kvec,\Qp,q_{z,l} \right)\), so pixels \(\mathbf{k}\) outside the double and triple overlap regions can be discarded when solving Eq.~\eqref{eq:g-5dctf}. We will denote the detector pixel coordinates inside the double or triple overlap regions, i.e. those for which \(A\left( \mathbf{k'} \right)\left\lbrack A\left( \mathbf{k' -}\Qp \right) + A\left( \mathbf{k' +}\Qp \right) \right\rbrack > 0,\) as \(\left\{ \mathbf{k}^{\mathbf{'}} \right\} = \mathbf{k}_{1}^{'},\ldots\mathbf{k}_{M^{'}\left( \Qp \right)}^{'}\) for each \(\Qp\). As the shape and area of the overlapping regions vary with \(\Qp\), so too do the number of detector pixel coordinates \(M^{'}\left( \Qp \right)\) therein.

The notation of the previous section seemed to imply a slice sampling matching that for converged multislice calculations in atomic resolution STEM, typically \(\Delta z\) \textasciitilde2 \AA{}. However, given the physics of information transfer in the tWPOA model, that would be a poor choice for the inverse problem. Eq.~\eqref{eq:qz-limit} gives the global maximum over all \(\Qp\) of the information limit along \(q_{z}\) to be \(\max{\left( q_{z,\max}\left( \Qp \right) \right) =}\lambda k_{0}^{2}/2\) as shown in Fig.~\ref{fig:1}f. Consequently, the Nyquist sampling rate for reconstructing the 3D structure sufficiently is \(\left( \lambda k_{0}^{2} \right)^{- 1}\). If the thickness range to solve the structure is set to be \(t^{'}\), the minimum number of slices is then \(N_{z} = \left\lceil t^{'}\lambda k_{0}^{2} \right\rceil\) with a slice thickness of \(\Delta z_{\mathrm{tWPOA}} = t^{'}/(N_{z} - 1)\). As per the practice in multislice ptychography \citep{chen2021}, it is preferable for the choice of \(t^{'}\) to appreciably exceed the actual sample thickness \(t\), with the reconstructed potential tending to zero outside the sample being a check on the validity of the reconstruction. Here, we denote the z coordinates of the \(j\)-th slice to solve by \(\left\{ z_{j}^{'} \right\}_{j = 1}^{N_{z}}\), where \(z_{j}^{'} = (j - 1)\Delta z_{\mathrm{tWPOA}} + z_{1}^{'}\). For example, with an accelerating voltage of 300 kV and a convergence semi-angle of 40 mrad, the information limit would be 0.0406 \AA{}\textsuperscript{-1} and the corresponding Nyquist sampling \(\Delta z_{\mathrm{tWPOA}} =\)1.2 nm. If we set \(t^{'} =\)10 nm then \(N_{z} = 9\) is a sufficient number of the slices, and finer depth variation than this is not recoverable (within the linear approximation) due to the finite depth-of-field of the STEM probe.

In principle, the \(q_{z}\) sampling could differ for different \(\Qp\) values since as per Eq.~\eqref{eq:qz-limit} the information limit differs, but we opt for the same sampling to facilitate using a 3D Fourier transform to reconstruct the scattering potential in real space. However, we use the bound of Eq.~\eqref{eq:qz-limit} to limit the number of \(q_{z}\) values solved for at each \(\Qp\) to those for which \(\left| q_{z,l}^{} \right| \leq q_{z,\max}\left( \Qp \right)\). These we denote \(\left\{ q_{z}^{'} \right\} = q_{z,1}^{'},\ldots,q_{z,N_{z}^{'}\left( \Qp \right)}^{'}\). The size reduction regarding \(\left\{ \mathbf{k} \right\}\ \)and \(\left\{ q_{z} \right\}\) coordinates using the 5D-CTF characteristics helps reduce computational costs for solving the equations.

With these sampling considerations in mind, we solve Eq.~\eqref{eq:g-5dctf} as follows. Separately for each spatial frequency \(\Qp\), we recast Eq.~\eqref{eq:g-5dctf} as the matrix-vector product
\begin{equation}
\mathbf{g}(\Qp)
=
\widehat{\mathbf{B}}(\Qp)\widehat{\boldsymbol{\Phi}}(\Qp),
\label{eq:linear-system}
\end{equation}
where \(\mathbf{g}\left(\Qp\right)\) is a measured-intensity (column) vector
\begin{equation}
\mathbf{g}(\Qp)
=
\begin{bmatrix}
G(\kvec'_1,\Qp)\\
\vdots\\
G\!\left(\kvec'_{M'(\Qp)},\Qp\right)
\end{bmatrix}
\in\mathbb{C}^{M'(\Qp)},
\label{eq:g-vector}
\end{equation}
\(\mathbf{B}\left(\Qp\right)\) is a 5D-CTF matrix
\begin{equation}
\begin{aligned}
\widehat{\mathbf{B}}(\Qp)
&=
\left[
\begin{smallmatrix}
\widehat{\beta}(\kvec'_1,\Qp,q'_{z,1})
& \cdots &
\widehat{\beta}\!\left(\kvec'_1,\Qp,q'_{z,N'_z(\Qp)}\right)
\\[-4pt]
\vdots & \ddots & \vdots
\\
\widehat{\beta}\!\left(\kvec'_{M'(\Qp)},\Qp,q'_{z,1}\right)
& \cdots &
\widehat{\beta}\!\left(\kvec'_{M'(\Qp)},\Qp,q'_{z,N'_z(\Qp)}\right)
\end{smallmatrix}
\right]
\\[-1pt]
&\in \mathbb{C}^{M'(\Qp)\times N'_z(\Qp)},
\end{aligned}
\label{eq:B-matrix}
\end{equation}
and \(\widehat{\boldsymbol{\Phi}}\left(\Qp\right)\) denotes a phase (column) vector
\begin{equation}
\widehat{\boldsymbol{\Phi}}(\Qp)
=
\begin{bmatrix}
\widehat{\Phi}_{q'_{z,1}}(\Qp)\\
\vdots\\
\widehat{\Phi}_{q'_{z,N'_z(\Qp)}}(\Qp)
\end{bmatrix}
\in\mathbb{C}^{N'_z(\Qp)}.
\label{eq:Phi-vector}
\end{equation}
By solving Eq.~\eqref{eq:linear-system} for \(\widehat{\boldsymbol{\Phi}}\left(\Qp\right)\) for each \(\Qp\) within the information limit for the lateral spatial frequency \(0 < \left| \Qp \right| \leq 2k_{0}\), the 3D Fourier-component distribution \(\widehat{\Phi}\left( \Qp,q_{z,l} \right)\) is obtained. A 3D Fourier transformation with respect to \(\Qp\) and \(q_{z,l}\) then gives the 3D structure in real space as schematically shown in Fig.~\ref{fig:1}d.

To show that Eq.~\eqref{eq:linear-system} is likely to be an overdetermined system, we evaluate \(M^{'}\left( \Qp \right)\), the number of discrete detector pixels \(\left\{ \mathbf{k'} \right\}\) within the double or triple overlap regions, and \(N_{z}^{'}\left( \Qp \right)\), the number of sampling in the frequency coordinate \(\left\{ q_{z}^{'} \right\}\), in a practical situation. Consider a square area just spanning the bright-field disk sampled by 40\(\times\)40 pixels, and assume an accelerating voltage of 300 kV and a probe forming aperture of 40 mrad. With these assumptions, the bright-field area comprises 1256 detector pixels and \(M^{'}\left( \Qp \right)\) monotonically decreases from this value with increasing \(\left| \Qp \right|\) as the scattered disks are shifted away from the transmitted disk (see Fig.~\ref{fig:1}e). As discussed above, \(N_{z}^{'}\left( \Qp \right)\) shows a parabolic-like distribution with maximum \(N_{z}^{'}\left( \Qp \right) = 9\) at \(\left| \Qp \right| = k_{0}\) and with even fewer sampling points at lower and higher \(\left| \Qp \right|\) values (see green lines in Fig.~\ref{fig:1}f). As shown in Fig.~\ref{fig:1}g, \(M^{'}\left( \Qp \right)\) is thus much larger than \(N_{z}^{'}\left( \Qp \right)\), except near the information limit along the lateral spatial frequency (\(\left| \Qp \right|\sim 2k_{0}\)), and Eq.~\eqref{eq:linear-system} can be solved via methods such as singular value decomposition (SVD).

\section{Methods}\label{methods}

\subsection{4D-STEM simulations}\label{d-stem-simulations}

We performed 4D-STEM simulations using in-house simulation codes employing the multislice model and linear scattering model, as described in Eqs.~\eqref{eq:multislice} and \eqref{eq:twpoa-wave}, respectively. We assumed monolithic SrTiO\textsubscript{3} (STO) single crystal and twisted bilayer STO structure for the simulation. The single crystal STO structure parameters are taken from \citep{kiat2000}. The bilayer STO structure was made by stacking rotated single crystal STO structures, with no structural reconstruction in the interface region. The slice thickness for the 4D-STEM simulations was 2 \AA{}, and an accelerating voltage was assumed to be 300 kV. Before the 3D LINV reconstruction, we binned the diffraction plane of the simulated 4D-STEM data and cropped the area just outside of the bright field, resulting in the 4D data size of 97 \(\times\) 97 \(\times\) 48\(\  \times\) 48 for a convergence semi-angle of 40 mrad, where the scan step was 0.123 \AA{} and the detector pixel size was 0.0854 \AA{}\textsuperscript{-1}.

\subsection[Experimental 4D-STEM acquisition on a ZrO2/Hf0.5Zr0.5O2 membrane]{Experimental 4D-STEM acquisition on a\protect\\ ZrO$_2$/Hf$_{0.5}$Zr$_{0.5}$O$_2$ membrane}\label{subsec:experiment}

We fabricated a STO/\allowbreak La\textsubscript{0.7}Sr\textsubscript{0.3}MnO\textsubscript{3} (LSMO)/\allowbreak ZrO\textsubscript{2} (ZO)/\allowbreak Hf\textsubscript{0.5}Zr\textsubscript{0.5}O\textsubscript{2} (HZO) thin film by means of pulsed laser deposition (PLD). The LSMO buffer layer was chemically etched and the ZO/HZO layer transferred onto a TEM grid with a carbon support film as a freestanding membrane, as detailed in the literature \citep{shen2025}. The thickness of each layer was 2.5 nm with the total membrane thickness being 5 nm. 4D-STEM measurement was performed using the Thermo Fisher Scientific Spectra $\varphi$ FEG TEM equipped with probe and image aberration correctors and an electron microscope pixel array detector (EMPAD), enabling 4D-STEM data acquisition with a readout time of 0.14 ms per probe position. We set the accelerating voltage and convergence semi angle to be 300 kV and 40 mrad, respectively. The detector has 128 \(\times\) 128 pixels and we adjusted the camera length so that the bright-field area is sampled by 80 \(\times\) 80 pixels, followed by binning by a factor of 2 after the data acquisition. Consequently, we obtained a 4D-STEM dataset with a size of 256 \(\times\) 256 \(\times\) 40\(\  \times\) 40, where the size of scanning points is 256 \(\times\) 256 and the area outside of the bright field has been cropped, similarly to the simulation described above. The probe current was set to 15 pA, resulting in an electron dose of 1.26\(\times\)10\textsuperscript{4} electrons per probe position or 1.87\(\times\)10\textsuperscript{5} e/\AA{}\textsuperscript{2} as per the scan step of 0.26 \AA{}.

\subsection{3D structure reconstruction}\label{d-structure-reconstruction}

After having binned and cropped the acquired simulated/\allowbreak experimental 4D-STEM dataset \(I\left( \kvec,\Rp \right)\), we Fourier transform it with respect to probe position to generate the 4D complex data \(G\left(\kvec,\Qp\right)\). At each spatial frequency coordinate \(\Qp\) within the lateral information limit (\(0 < \left| \Qp \right| \leq 2k_{0}\)), we construct the matrix-based equation shown in Eq.~\eqref{eq:linear-system} as per Eqs.~\eqref{eq:g-vector} and \eqref{eq:B-matrix}, and solve the equation for \(\widehat{\boldsymbol{\Phi}}\left(\Qp\right)\) to obtain the Fourier components of the 3D structure \(\widehat{\Phi}\left( \Qp,q_{z,l} \right)\). 3D Fourier transformation then gives the 3D structure in real space. The computing time is modest: as detailed in Supplementary Note 1, in our NumPy-based \citep{harris2020} implementation (i.e., without seeking the speed-up doubtless possible with modern graphical processing unit (GPU) computation) the computing time of the reconstructions shown varied with data size between 14 and 134 seconds.

\section{Results}\label{results}

\begin{figure*}[t!]
\centering
\includegraphics[width=0.85\textwidth]{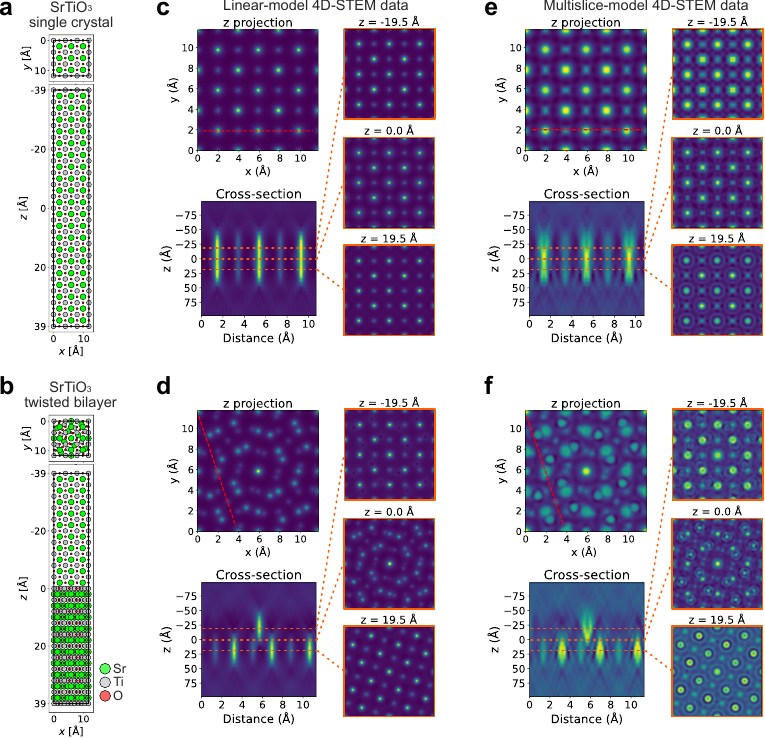}
\caption{Atomic structure models of (a) single-crystal SrTiO\textsubscript{3} [001] and (b) bilayer SrTiO\textsubscript{3}. From simulated 4D-STEM datasets assuming the tWPOA model for the structures shown in (a) and (b), we reconstructed the 3D phase distributions shown in (c) and (d), respectively. In each, z-projection, cross-section slice, and representative plan-view slices are shown. From simulated 4D-STEM datasets assuming the multislice model, we similarly reconstructed the 3D structures shown in (e) and (f). All simulations assume an accelerating voltage of 300 kV and a convergence semi-angle of 40 mrad.}
\label{fig:2}
\end{figure*}

\begin{figure*}[t!]
\centering
\includegraphics[width=0.9\textwidth]{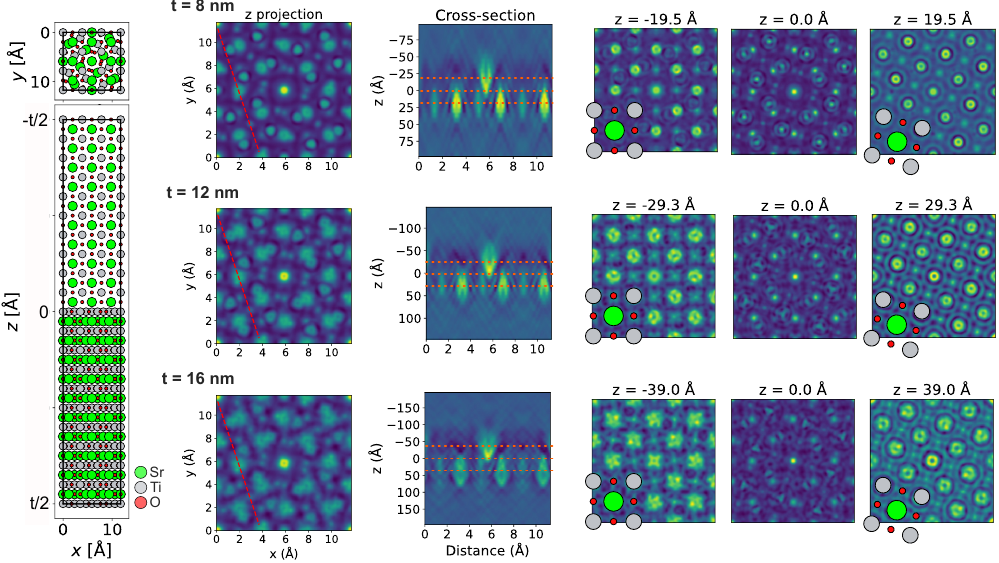}
\caption{Structure reconstructions for stacked bilayer SrTiO\textsubscript{3} for various total specimen thicknesses \(t\) with the thickness of each layer being \(t/2\). An accelerating voltage of 300 kV and a convergence semi-angle of 40 mrad are assumed. The cross-section profiles are taken from the red dashed lines in the z projections. The z coordinates for extracting the representative plan-view slices are highlighted by orange dashed lines in the cross-section profiles. Atomic structure models are overlaid on the extracted plan-view slices.}
\label{fig:3}
\end{figure*}

\subsection{Linear-model-based reconstruction in the presence of multiple scattering}\label{linear-model-based-reconstruction-in-the-presence-of-multiple-scattering}

In this section, we explore the quality of the tWPOA-based 3D structure reconstruction from 4D-STEM data by applying it to two different forward scattering models: the linear scattering model of Eq.~\eqref{eq:twpoa-wave} (where it should only be limited by the probe depth of field) and the multislice model of Eq.~\eqref{eq:multislice} (where there will be additional model mismatch due to dynamical scattering). As a target crystalline structure, we assume a monolithic single crystal of STO along {[}001{]} zone axis (Fig.~\ref{fig:2}a) and a twisted bilayer structure (Fig.~\ref{fig:2}b) with the lower layer rotated by \(\arctan(1/3)\) (which corresponds to a $\Sigma 3$ grain boundary and has a relatively small repeat distance). For both structures, the thickness \(t\) is 7.8 nm (20 unit cells) and the focal plane is located in the middle of the specimen (i.e., \(z_{1} = - t/2 = - 3.9\) nm), which, as detailed in Supplementary Note 5, gives better reconstruction accuracy than a surface-focused or overfocused probe.

We first reconstruct the 3D structure of single-crystal STO from tWPOA-based 4D-STEM data as shown in Fig.~\ref{fig:2}c. From the cross-section (x-z) slices and representative plan-view (x-y) plane slices, the reconstructed phase map shows the atomic structure of STO well. For the reconstruction, we set \(t'\)=19.5 nm (with the number of slices \(N_{z}\)=17), which given defocus to the midplane should result in vacuum layers with a thickness of \(0.75t\) either side of the specimen. In this condition, the cross-section slice through the reconstructed result shows that the phase distribution corresponds well to the actual thickness.

We next reconstruct the 3D structure of twisted bilayer STO structure from a tWPOA-based 4D-STEM data as shown in Fig.~\ref{fig:2}d. Whereas the upper part of the reconstructed 3D structure (\(z <\) 0 \AA{}) is identical to what has been observed in Fig.~\ref{fig:2}c, the lower part shows the atomic structure of the STO lattice twisted from the upper part. We note that, as discussed in the Theoretical Framework section, the reconstructed depth resolution is band-width limited due to the 5D-CTF nature, leading to an elongation along the z direction of features in the reconstruction. These results show that our method can retrieve 3D structure as an inversion problem in the linear scattering model.

We next reconstruct 3D structure from the multislice-simulated 4D-STEM data for the single crystal and twisted bilayer structure of STO. For single crystal STO, the 3D reconstruction in Fig.~\ref{fig:2}e shows contrast corresponding to the atomic structure. However, unlike the linear-model case, each atomic-site contrast shows artefacts. In particular, the heavier atomic sites (Sr and TiO) have a doughnut-shape contrast, which is commonly observed in the images reconstructed from methods based on the (weak) phase object approximation \citep{yang2017,ooe2021}. This indicates that the multiple scattering, which is not accounted for in the inversion process, affects the reconstruction result. Nevertheless, both heavy and light atomic sites are evident, as is the approximate thickness range of the sample. For the twisted bilayer STO structure, shown in Fig.~\ref{fig:2}f, the reconstruction again shows doughnut-shape artifacts particularly at the heavier atomic columns. However, both heavy and light elements in the atomic structure in the two different layers are resolved along the depth direction with comparable clarity to the linear-model case in Fig.~\ref{fig:2}d.

Since multiple scattering increases with sample thickness, we would expect the model mismatch between multislice simulations and the tWPOA model to increase with sample thickness. To explore this, we prepared twisted bilayer STO atomic structures with thicknesses of \(t\) = 8 nm (same as Fig.~\ref{fig:2}f), 12 nm, and 16 nm, where the lower part with thickness of \(t/2\) is twisted from the upper part by the same angle as earlier. The reconstructions in Fig.~\ref{fig:3} show more evident ring-shaped artefacts and lower contrast, particularly in thicker specimens. Nevertheless, the twisted atomic structures are resolved even with a thickness of 16 nm, which is quite ``thick'' by the standards of linear scattering approximations \citep{ooe2021}. In addition, the layer interface and the specimen thickness along the \emph{z} direction remain evident. Overall, the structures are reasonably well resolved in 3D despite the stronger multiple scattering effect.

The information limit along the z direction (see Eq.~\eqref{eq:qz-limit} and Fig.~\ref{fig:1}f) depends on the probe convergence angle. Figs.~\ref{fig:2} and \ref{fig:3} assume a 40 mrad convergence angle. Supplementary Note 4 shows that the depth resolution would be reduced at 30 mrad but increased at 50 mrad. Moreover, reconstruction artefacts tend to reduce with increasing convergence angle since plane wave components with the larger tilt angles away from the exact zone axis condition tend to be less affected by multiple scattering \citep{brown2019}.

Though not accounting for multiple scattering nor reaching the resolution achieved by multislice ptychography \citep{chen2021}, these reconstructions are fast: the 3D reconstructions shown here took a few tens of seconds on a workstation CPU, not GPU (see Supplementary Note 1). Further, the tWPOA-based reconstruction\textquotesingle s direct (cf. iterative) inversion scheme has minimal hyperparameter tuning and/or explicit regularisation, providing deterministic 3D reconstruction. We therefore think this method promising for near-real-time analysis of atomic structures inhomogeneous along the depth direction. Structure and thickness estimated from this method could also inform subsequent and further structure refinement via multislice ptychography.

\subsection{Effects of partial coherence}\label{effects-of-partial-coherence}

\begin{figure*}[t!]
\centering
\includegraphics[width=0.85\textwidth]{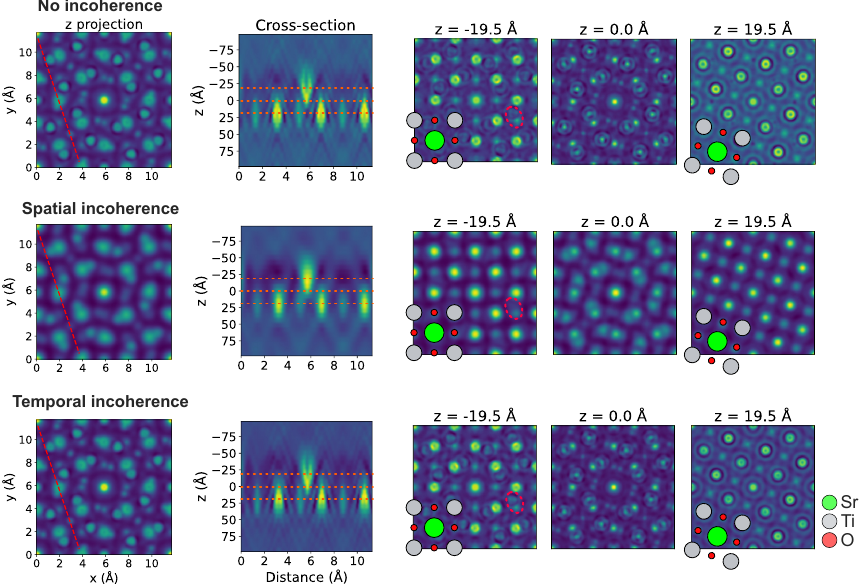}
\caption{Comparison between 3D reconstruction results for perfect coherence (top row), spatial incoherence with a source size of 0.6 \AA{} in FWHM (middle row), and temporal incoherence with a chromatic aberration coefficient of 1.2 mm and an energy spread of 0.7 eV in FWHM (bottom row). An accelerating voltage of 300 kV, a convergence semi-angle of 40 mrad, and a specimen thickness of 7.8 nm are assumed.}
\label{fig:4}
\end{figure*}

We next investigate the effect of partial coherence, which is always inherent in experimental 4D-STEM measurements. For spatial incoherence, we assume a Gaussian distribution for the finite effective source size \citep{verbeeck2012,nguyen2014}:
\begin{equation}
I_{\mathrm{incoh}}^{\mathrm{spatial}}(\kvec,\Rp)
=
I(\kvec,\Rp)
\otimes_{\Rp}
\frac{1}{2\rmpi\sigma_s^2}
\exp\!\left(-\frac{|\Rp|^2}{2\sigma_s^2}\right).
\label{eq:spatial-incoh}
\end{equation}
Temporal incoherence accounts for the defocus spread due to the energy fluctuation and finite chromatic aberration coefficient \citep{ishikawa2016,oxley2020}. We assume a Gaussian distribution for the source energy spread distribution:
\begin{equation}
I_{\mathrm{incoh}}^{\mathrm{temporal}}(\kvec,\Rp)
=
I(\kvec,\Rp;z_1+t)
\otimes_{z_1}
\frac{1}{\sqrt{2\rmpi}\,\sigma_t}
\exp\!\left(-\frac{|z_1|^2}{2\sigma_t^2}\right).
\label{eq:temporal-incoh}
\end{equation}
While in practice both effects are present together, we use simulation to explore them separately to elucidate their distinct impact on 3D LINV reconstructions.

The results of the 3D LINV reconstruction for multislice-simulated 4D-STEM datasets from 7.8-nm-thick twisted bilayer SrTiO\textsubscript{3} in the presence of spatial/temporal incoherence are shown in Fig.~\ref{fig:4}. For spatial incoherence, we assume an effective source size of 0.6 \AA{} full-width-half-maximum (FWHM). For temporal incoherence, we assume energy spread of \(\Delta E\)=0.7 eV in FWHM and a chromatic aberration coefficient of \(C_{c} =\)1.2 mm, giving a defocus spread of \(\sigma_{t} = C_{c}\frac{\Delta E}{E}/2\sqrt{2\ln 2} =\)11.9 \AA{}. The depth-of-focus obtained under the assumed condition corresponds to \(2\left( \lambda/\alpha^{2} \right)\) = 24.6 \AA{}, compared to which the defocus spread due to chromatic aberration is not negligible and so may worsen the depth resolution as discussed in the literature \citep{ishikawa2016}.

As discussed in the previous section, reconstructed 3D phase contrast tends to have doughnut-shaped artefacts above heavier atomic columns due to multiple scattering. In the presence of spatial incoherence, however, the doughnut-shapes are blurred out, and even heavier atomic columns are visualised without those artefacts. Though this does not fix the artefacts originating from the model mismatch, it aids visual interpretation of atomic configurations in the reconstructed results. Temporal incoherence does not, in the lateral dimension, affect the reconstruction result to nearly the same degree as spatial incoherence does. However, with temporal incoherence the extracted slice at z=-19.5 \AA{} shows atomic-column contrast contributions from both layers, as highlighted by a red-dashed circle in Fig.~\ref{fig:4}. Hence, as investigated in other depth-sectioning techniques in STEM \citep{ishikawa2016}, temporal incoherence due to energy spread and chromatic aberration worsens the depth resolution.

\subsection{Effects of finite dose and noise normalisation}\label{effects-of-finite-dose-and-noise-normalisation}

\begin{figure*}[t!]
\centering
\includegraphics[width=0.9\textwidth]{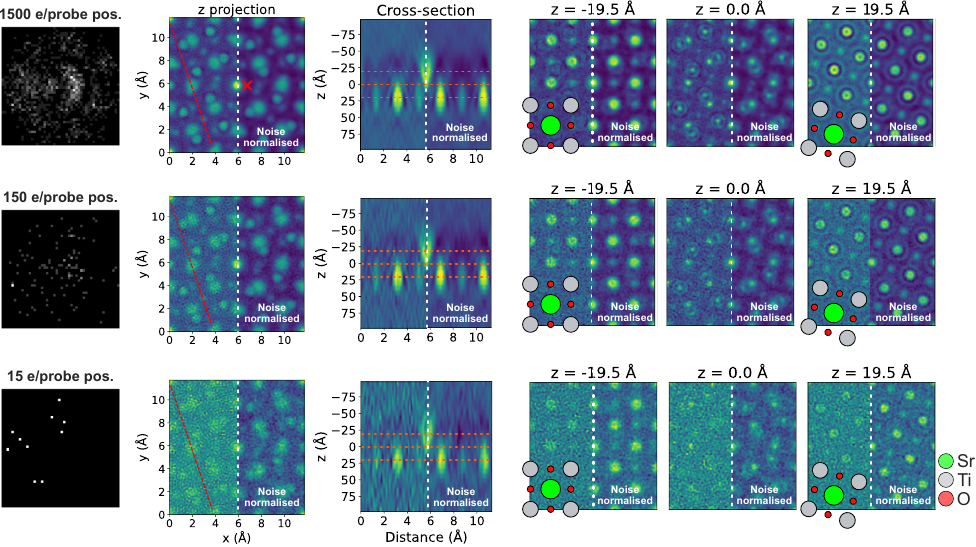}
\caption{3D reconstruction results for a 7.8-nm-thick stacked bilayer SrTiO\textsubscript{3} structure across a series of finite doses. For each assumed dose, reconstruction results are shown as z-projections, cross-section slices, and extracted plan-view slices, without and with noise normalisation (in the left- and right-side of each panel, respectively). The red cross in the z-projection reconstructed from the 1500 e/probe-position dataset indicates the probe position from which the diffraction patterns shown in the left-most column were extracted for each dose condition. Red lines in the z projections show the area for taking the cross-sections, and orange lines in the cross sections correspond to z coordinates to extract the plan-view slices.}
\label{fig:5}
\end{figure*}

We next consider effects on the 3D-LINV reconstruction of shot noise originating from finite electron doses. We first add noise to simulated 4D-STEM data by assuming Poisson counting statistics, and then reconstruct the 3D structure as shown in left-side panels of Fig.~\ref{fig:5}. Under lower dose conditions, the reconstructed results have worse atomic-site visibility, which manifests in the \emph{z}-projection, cross-section slices, and plan-view slices of the reconstructed phase. In particular, the lower-dose results have substantial high-frequency noise fluctuation.

In (projected) two-dimensional phase images reconstructed from 4D-STEM data, such as single side band (SSB) ptychography, optimum bright-field (OBF), and tilt-corrected bright field (tcBF) techniques, the transfer characteristics not only of contrast but also of noise have been investigated \citep{seki2018b,oleary2021,ooe2025,varnavides2026,bennemann2026,ma2026}. By normalising the contrast transfer by the noise transfer as a function of spatial frequency, the SNR at each frequency coordinate can be calculated. The result is called a noise-normalised CTF, SNR transfer function (SNRTF), spectral SNR (SSNR), or detective quantum efficiency (DQE) in that literature and offers a fairer metric for comparing different imaging methods under the finite-dose conditions. Furthermore, the noise-normalisation of a reconstructed phase image provides better visualisation of atomic structure by avoiding the noise amplification at specific spatial frequency domains \citep{oleary2021,ooe2023}.

\begin{figure}[!htbp]
\centering
\includegraphics[width=1.01\columnwidth]{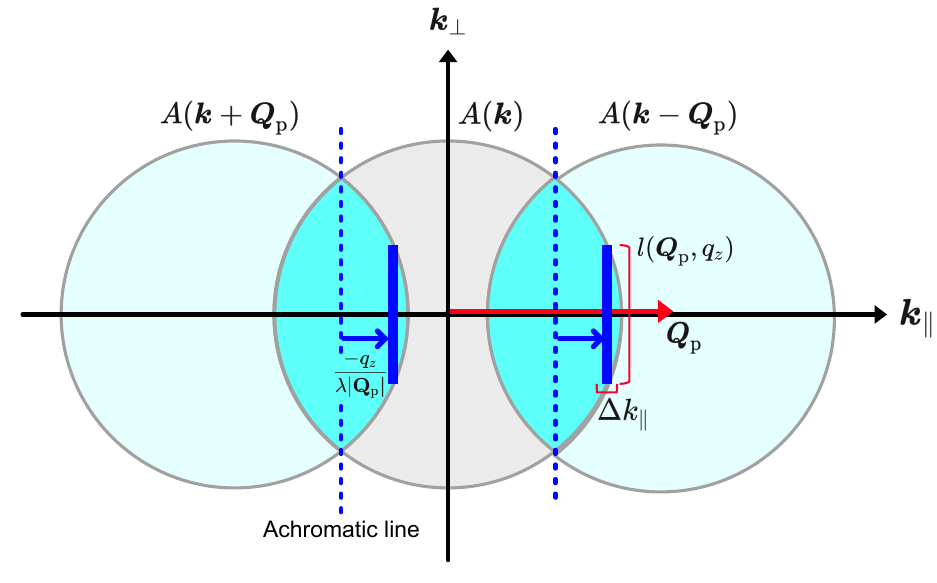}
\caption{Schematic illustration of a geometrical configuration of transmitted disk, scattered disks, and the lines corresponding to the possible detector coordinates with non-zero 5D-CTF for given \(\Qp\) and \(q_{z,l}\). Here, the axes of coordinates are shown as \(\mathbf{k}_{\bot} = k_{\bot}{\widehat{\mathbf{e}}}_{\mathbf{\bot}}\) and \(\mathbf{k}_{\parallel} = k_{\parallel}{\widehat{\mathbf{e}}}_{\mathbf{\parallel}}\) where \({\widehat{\mathbf{e}}}_{\mathbf{\bot}}\) and \({\widehat{\mathbf{e}}}_{\mathbf{\parallel}}\) are perpendicular and parallel to \(\Qp\), respectively. The red arrow indicates the \(+\Qp\) vector and the solid blue lines show the area with non-zero 5D-CTF (with width \(\mathrm{\Delta}k_{\parallel}\) and height \(l\left( \Qp,q_{z} \right)\) as discussed in Supplementary Note 3), which are shifted from the achromatic lines by \(- q_{z,l}/\lambda\left| \Qp \right|\) within the double overlap regions.}
\label{fig:6}
\end{figure}

The noise transfer as a function of 3D spatial frequency coordinate can be evaluated as follows. In the 5D CTF shown in Eq.~\eqref{eq:5dctf}, let us assume for simplicity that the sampling rates of \(\mathbf{k}\), \(\Qp\), and \(q_{z,l}\) are commensurate so that \(W\left( q_{z,l} - \omega_{\pm} \right)\) reduces to a Kronecker delta function \(\delta\left( q_{z,l} - \omega_{\pm} \right)\), yielding
\begin{equation}
\begin{aligned}
\widehat{\beta}(\kvec,\Qp,q_{z,l})
={}&\ii A(\kvec)\Bigl[
A(\kvec-\Qp)\delta(q_{z,l}-\omega_-)
\\[-1pt]
&\hspace{4.2em}-A(\kvec+\Qp)\delta(q_{z,l}-\omega_+)
\Bigr].
\end{aligned}
\label{eq:5dctf-delta}
\end{equation}
In addition to the double/triple overlap constraint discussed in Fig.~\ref{fig:1}e, \(\widehat{\beta}\left( \kvec,\Qp,q_{z,l} \right)\) will only have non-zero values if
\begin{equation}
q_{z,l}
=
\omega_{\pm}
=
-\lambda\left(
\kvec\cdot\Qp
\pm
\frac{|\Qp|^2}{2}
\right).
\label{eq:qz-omega}
\end{equation}
This induces another constraint to the \(\left\{ \mathbf{k} \right\}\) coordinates with non-zero 5D-CTF at a fixed spatial frequency coordinate \(\left( \Qp,q_{z,l}\mathbf{\ } \right)\), which takes the form of lines normal to the \(\Qp\) vector within the overlap regions as shown in Fig.~\ref{fig:6}. Here, we denote the detector plane coordinate as \(\mathbf{k =}\mathbf{k}_{\mathbf{\bot}}\mathbf{+}\mathbf{k}_{\mathbf{\parallel}}\mathbf{=}k_{\bot}{\widehat{\mathbf{e}}}_{\mathbf{\bot}}\mathbf{+}k_{\parallel}{\widehat{\mathbf{e}}}_{\mathbf{\parallel}}\), where \({\widehat{\mathbf{e}}}_{\mathbf{\bot}}\) and \({\widehat{\mathbf{e}}}_{\mathbf{\parallel}}\) are unit vectors perpendicular and parallel to \(\Qp\) vector. With this notation, Eq.~\eqref{eq:qz-omega} simplifies to
\begin{equation}
k_{\parallel}
=
\begin{cases}
\displaystyle
\frac{|\Qp|}{2}
-
\frac{q_{z,l}}{\lambda|\Qp|},
& \text{for } +\Qp \text{ shifted disk},\\[8pt]
\displaystyle
-\frac{|\Qp|}{2}
-
\frac{q_{z,l}}{\lambda|\Qp|},
& \text{for } -\Qp \text{ shifted disk}.
\end{cases}
\label{eq:kparallel}
\end{equation}
Eq.~\eqref{eq:kparallel} indicates that, with \(q_{z,l} = 0\), the \(\left\{ \mathbf{k} \right\}\) coordinates with non-zero 5D-CTF must have \(k_{\parallel} = \pm \frac{\left| \Qp \right|}{2}\), which corresponds to the so-called achromatic lines \citep{pennycook2015,pennycook2019} at the midpoint of the intersection of the transmitted and scattered disks, shown as dashed blue lines in Fig.~\ref{fig:6}. For non-zero \(q_{z,l}\) values, this line shifts by \(- \frac{q_{z,l}}{\lambda\left| \Qp \right|}\) along the \(\Qp\) vector, shown as thick blue lines in Fig.~\ref{fig:6}. The detector coordinates on this line can have non-zero transfer from the 3D potential to the 4D-STEM data at fixed \(\Qp\) and \(q_{z,l}\). Hence, by analogous reasoning to that used for the contrast and noise transfer characteristics in SSB ptychography, the noise transfer function \(N\left( \Qp,q_{z,l} \right)\) of the reconstructed 3D phase is deduced to be
\begin{equation}
\begin{aligned}
N(\Qp,q_{z,l})
\propto{}&
\sqrt{\lambda|\Qp|}
\\[-1pt]
&\times
\left[
 k_0^2-
 \left(
  \frac{|\Qp|}{2}
  +
  \frac{|q_z|}{\lambda|\Qp|}
 \right)^2
\right]^{-1/4}.
\end{aligned}
\label{eq:noise-transfer}
\end{equation}
The details of the noise transfer function derivation are given in Supplementary Note 3. Normalising the noise level as per Eq.~\eqref{eq:noise-transfer}, the reconstructed 3D phase image will have a uniform-amplitude noise distribution with improved atomic visibility under finite dose conditions.

We show the noise-normalised 3D reconstruction results in the right-hand-side of each panel in Fig.~\ref{fig:5}. Compared with the reconstructions without noise normalisation, the normalised result shows better visualisation of atomic structures even under lower dose conditions with no appreciable change to the depth resolution. Noise-normalisation improves the visibility and applicability of our 3D structure reconstruction approach under practical conditions.

\begin{figure*}[h]
\centering
\includegraphics[width=0.9\textwidth]{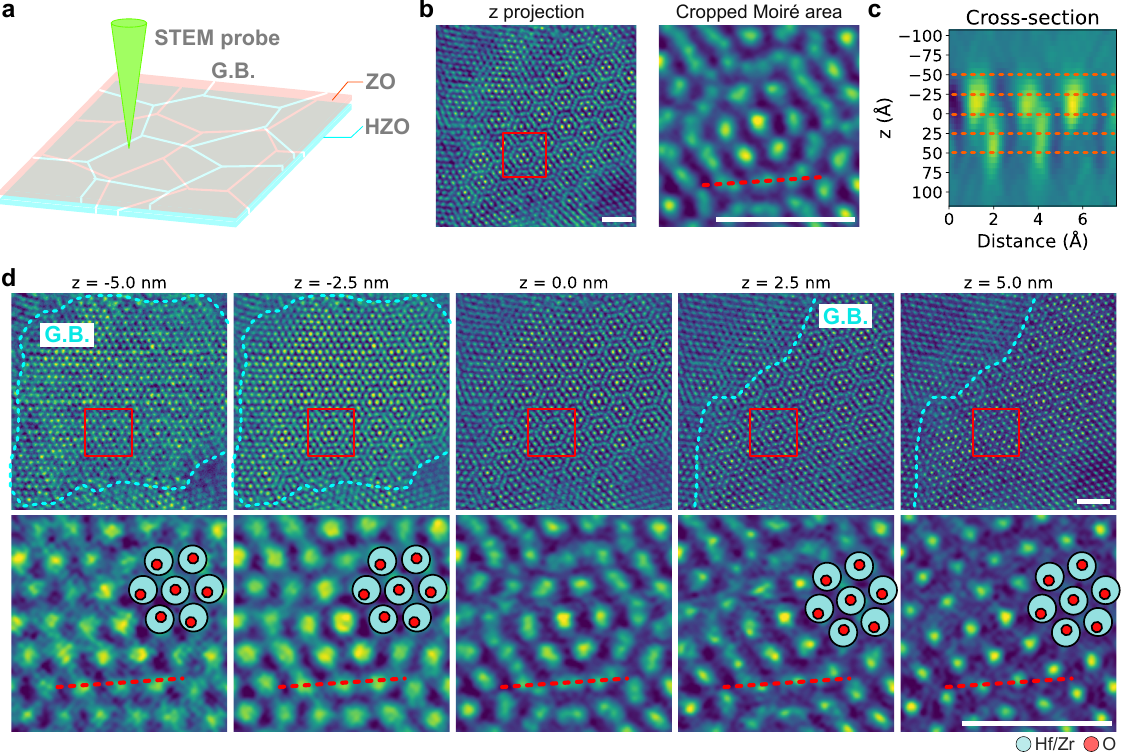}
\caption{(a) Schematic of the ZO/HZO freestanding membrane specimen. Each layer comprises polycrystalline textures with nanocrystals and grain boundaries. (b) \emph{z}-projection and its cropped magnification from the red box area highlighting a Moir\'e region in the z projection. (c) Cross-section slice obtained from a red-dashed line in the magnified image with Moir\'e area. (d) Five representative plan-view slices with the z coordinates highlighted by dotted orange lines in (c). Upper row shows the whole field of view with different z coordinates and lower row shows their digital zoom from the red-box areas. Grain boundaries are highlighted by cyan-dashed lines. Atomic structure models are overlaid. All scale bars indicate 1 nm.}
\label{fig:7}
\end{figure*}

\subsection{Experimental demonstration}\label{experimental-demonstration}
We finally demonstrate a 3D reconstruction from an experimental 4D-STEM dataset of a freestanding ZrO\textsubscript{2} (ZO)/Hf\textsubscript{0.5}Zr\textsubscript{0.5}O\textsubscript{2} (HZO) membrane (see the Methods section for details of the specimen and the 4D-STEM measurement). As reported in our previous study \citep{shen2025} and depicted in Figure~\ref{fig:7}a, each ZO and HZO layer has a polycrystalline texture comprising nanocrystals and grain boundaries, resulting in complexed atomic structures as a projection. In this study, we acquired a 4D-STEM dataset from an area with a Moir\'e structure and performed 3D structure reconstruction with the noise-normalisation. We note that, for the present 4D data, the reconstruction duration was approximately 2 minutes using a standard workstation (see the Supplementary Note 1), fast enough to allow on-the-spot evaluation of the retrieved 3D structure during the STEM operation.

Fig.~\ref{fig:7}b shows a z-projection of the reconstructed 3D phase, visualising Moir\'e patterns in the middle of the field of view. In the Moir\'e area, the cross-section slice shown in Fig.~\ref{fig:7}c clearly resolves the two layers along the depth direction. Fig.~\ref{fig:7}d shows five plan-view slices extracted from the reconstructed result. The upper slices show a nanocrystal surrounded by a grain boundary (G.B.) in the middle of the field of view. This structure corresponds to a tetragonal phase normal to the (101) plane in fluorite-type crystals. The slice in the middle depth shows a Moir\'e structure similar to what was observed in the z-projection. In the lower part, another tetragonal nanocrystal with different in-plane orientation is visualised. In addition, the lower part has a grain boundary in a different area from the upper layer. This result shows that our 3D-LINV STEM method successfully resolved an atomic structure nonuniform along the depth direction from experimental 4D-STEM data.

\section{Concluding Remarks}\label{sec:conclusion}

We have developed 3D-LINV STEM, a linear-inversion technique for reconstructing 3D structure from a single-defocus 4D-STEM dataset. We formalised a linear equation relating the measured 4D-STEM intensity and the unknown 3D phase distribution within the thick weak phase object approximation, and have shown that the equation can be solved directly, without explicit regularisation or computationally expensive iterative calculations, such that it could be performed on-the-fly during the timeframe for experimental STEM data acquisition. We demonstrated that the algorithm effectively reconstructs atomic structures that are inhomogeneous along the depth direction, at atomic lateral resolution and nanometre depth resolution and in the presence of dynamical scattering, across a variety of conditions such as a sample thickness, convergence semi-angle, partial coherence, and finite dose. The 3D reconstruction demonstrated from experimental 4D-STEM data of a freestanding ZO/HZO stacked membrane specimen shows the capability of our technique for 3D structure reconstruction.

\section{Availability of Data and Materials}\label{availability-of-data-and-materials}

The Python-based 3D-LINV STEM scripts and example 4D-STEM datasets are available at: \url{https://github.com/KouOoe/3d-linv-stem}.

\section{Financial Support}\label{financial-support}

This research was supported under JSPS KAKENHI (Grant No. 22KJ3209, 23K13553, 24H01190, and 26H02199). This research was partially funded by the Australian Government through the Australian Research Council (project number DP250100935). The work was also supported by the Japan Science and Technology Agency (JST) as part of PRESTO, Grant No. JPMJPR24H3, and Adopting Sustainable Partnerships for Innovative Research Ecosystem (ASPIRE), Grants No. JPMJAP2312 and No. JPMJAP2314. The authors acknowledge the use of the instruments and scientific and technical assistance at the Monash Centre for Electron Microscopy, Monash University, a Microscopy Australia (ROR: 042mm0k03) facility supported by NCRIS. This research used equipment funded by ARC grants LE0454166 and LE17010118.


\section{Conflict of interest}
The authors declare that they have no competing interest.

\FloatBarrier
\bibliographystyle{plainnat}
\bibliography{reference}

@article{batson2002,
  author  = {Batson, P. E. and Dellby, N. and Krivanek, O. L.},
  title   = {Sub-{\AA}ngstrom resolution using aberration corrected electron optics},
  journal = {Nature},
  year    = {2002},
  volume  = {418},
  pages   = {617--620}
}

@article{bennemann2026,
  author  = {Bennemann, F. and Kirkland, A. I. and Muller, D. A. and Nellist, P. D.},
  title   = {Detective quantum efficiency-based comparison of {HRTEM} and ptychography phase imaging},
  journal = {Microscopy and Microanalysis},
  year    = {2026},
  volume  = {32},
  pages   = {ozag018}
}

@article{borisevich2006,
  author  = {Borisevich, A. Y. and Lupini, A. R. and Pennycook, S. J.},
  title   = {Depth sectioning with the aberration-corrected scanning transmission electron microscope},
  journal = {Proceedings of the National Academy of Sciences},
  year    = {2006},
  volume  = {103},
  pages   = {3044--3048}
}

@article{brown2019,
  author  = {Brown, H. G. and Ishikawa, R. and S{\'a}nchez-Santolino, G. and Shibata, N. and Ikuhara, Y. and Allen, L. J. and Findlay, S. D.},
  title   = {Large angle illumination enabling accurate structure reconstruction from thick samples in scanning transmission electron microscopy},
  journal = {Ultramicroscopy},
  year    = {2019},
  volume  = {197},
  pages   = {112--121}
}

@article{brown2022,
  author  = {Brown, H. G. and Pelz, P. M. and Hsu, S.-L. and Zhang, Z. and Ramesh, R. and Inzani, K. and Sheridan, E. and Griffin, S. M. and Schloz, M. and Pekin, T. C. and Koch, C. T. and Findlay, S. D. and Allen, L. J. and Scott, M. C. and Ophus, C. and Ciston, J.},
  title   = {A three-dimensional reconstruction algorithm for scanning transmission electron microscopy data from a single sample orientation},
  journal = {Microscopy and Microanalysis},
  year    = {2022},
  volume  = {28},
  pages   = {1632--1640}
}

@article{chen2021,
  author  = {Chen, Z. and Jiang, Y. and Shao, Y.-T. and Holtz, M. E. and Odstr{\v{c}}il, M. and Guizar-Sicairos, M. and Hanke, I. and Ganschow, S. and Schlom, D. G. and Muller, D. A.},
  title   = {Electron ptychography achieves atomic-resolution limits set by lattice vibrations},
  journal = {Science},
  year    = {2021},
  volume  = {372},
  pages   = {826--831}
}

@article{esser2023,
  author  = {Esser, B. D. and Etheridge, J.},
  title   = {Complementary {ADF-STEM}: a flexible approach to quantitative {4D-STEM}},
  journal = {Ultramicroscopy},
  year    = {2023},
  volume  = {243},
  pages   = {113627}
}

@article{gao2017,
  author  = {Gao, S. and Wang, P. and Zhang, F. and Martinez, G. T. and Nellist, P. D. and Pan, X. and Kirkland, A. I.},
  title   = {Electron ptychographic microscopy for three-dimensional imaging},
  journal = {Nature Communications},
  year    = {2017},
  volume  = {8},
  pages   = {163}
}

@article{haider1998,
  author  = {Haider, M. and Uhlemann, S. and Schwan, E. and Rose, H. and Kabius, B. and Urban, K.},
  title   = {Electron microscopy image enhanced},
  journal = {Nature},
  year    = {1998},
  volume  = {392},
  pages   = {768--769}
}

@article{harris2020,
  author  = {Harris, C. R. and Millman, K. J. and van der Walt, S. J. and Gommers, R. and Virtanen, P. and Cournapeau, D. and Wieser, E. and Taylor, J. and Berg, S. and Smith, N. J. and Kern, R. and Picus, M. and Hoyer, S. and van Kerkwijk, M. H. and Brett, M. and Haldane, A. and del R{\'i}o, J. F. and Wiebe, M. and Peterson, P. and G{\'e}rard-Marchant, P. and Sheppard, K. and Reddy, T. and Weckesser, W. and Abbasi, H. and Gohlke, C. and Oliphant, T. E.},
  title   = {Array programming with {NumPy}},
  journal = {Nature},
  year    = {2020},
  volume  = {585},
  pages   = {357--362}
}

@article{ishikawa2016,
  author  = {Ishikawa, R. and Pennycook, S. J. and Lupini, A. R. and Findlay, S. D. and Shibata, N. and Ikuhara, Y.},
  title   = {Single atom visibility in {STEM} optical depth sectioning},
  journal = {Applied Physics Letters},
  year    = {2016},
  volume  = {109},
  pages   = {163102}
}

@article{ishikawa2020,
  author  = {Ishikawa, R. and Shibata, N. and Taniguchi, T. and Ikuhara, Y.},
  title   = {Three-dimensional imaging of a single dopant in a crystal},
  journal = {Physical Review Applied},
  year    = {2020},
  volume  = {13},
  pages   = {034064}
}

@article{jeong2021,
  author  = {Jeong, J. and Cautaerts, N. and Dehm, G. and Liebscher, C. H.},
  title   = {Automated crystal orientation mapping by precession electron diffraction-assisted four-dimensional scanning transmission electron microscopy using a scintillator-based {CMOS} detector},
  journal = {Microscopy and Microanalysis},
  year    = {2021},
  volume  = {27},
  pages   = {1102--1112}
}

@article{kiat2000,
  author  = {Kiat, J.-M. and Baldinozzi, G. and Dunlop, M. and Malibert, C. and Dkhil, B. and M{\'e}noret, C. and Masson, O. and Fernandez-Diaz, M.-T.},
  title   = {Anharmonicity and disorder in simple and complex perovskites: a high energy synchrotron and hot neutron diffraction study},
  journal = {Journal of Physics: Condensed Matter},
  year    = {2000},
  volume  = {12},
  pages   = {8411--8425}
}

@article{kim2025,
  author  = {Kim, M.-S. and Lee, K. and Ishikawa, R. and Song, K. and Shahed, N. A. and Eom, K.-T. and Rzchowski, M. S. and Tsymbal, E. Y. and Shibata, N. and Mizoguchi, T. and Eom, C.-B. and Choi, S.-Y.},
  title   = {Charge disproportionation at twisted {SrTiO3} bilayer interface driven by local atomic registry},
  journal = {ACS Nano},
  year    = {2025},
  volume  = {19},
  pages   = {39714--39724}
}

@article{kimoto2011,
  author  = {Kimoto, K. and Ishizuka, K.},
  title   = {Spatially resolved diffractometry with atomic-column resolution},
  journal = {Ultramicroscopy},
  year    = {2011},
  volume  = {111},
  pages   = {1111--1116}
}

@article{kp2025,
  author  = {KP, H. and Xu, R. and Patel, K. and Crust, K. J. and Khandelwal, A. and Zhang, C. and Prosandeev, S. and Zhou, H. and Shao, Y. T. and Bellaiche, L. and Hwang, H. Y. and Muller, D. A.},
  title   = {Electron ptychography reveals a ferroelectricity dominated by anion displacements},
  journal = {Nature Materials},
  year    = {2025},
  volume  = {24},
  pages   = {1433--1440}
}

@article{krajnak2016,
  author  = {Krajnak, M. and McGrouther, D. and Maneuski, D. and O'Shea, V. and McVitie, S.},
  title   = {Pixelated detectors and improved efficiency for magnetic imaging in {STEM} differential phase contrast},
  journal = {Ultramicroscopy},
  year    = {2016},
  volume  = {165},
  pages   = {42--50}
}

@article{lee2025,
  author  = {Lee, C.-H. and Zeltmann, S. E. and Yoon, D. and Ma, D. and Muller, D. A.},
  title   = {{PtyRAD}: A high-performance and flexible ptychographic reconstruction framework with automatic differentiation},
  journal = {Microscopy and Microanalysis},
  year    = {2025},
  volume  = {31},
  pages   = {ozaf070}
}

@article{ma2026,
  author  = {Ma, D. and Li, G. and Muller, D. A. and Zeltmann, S. E.},
  title   = {Information in {4D-STEM}: where it is, and how to use it},
  journal = {Ultramicroscopy},
  year    = {2026},
  volume  = {283},
  pages   = {114351}
}

@article{maiden2009,
  author  = {Maiden, A. M. and Rodenburg, J. M.},
  title   = {An improved ptychographical phase retrieval algorithm for diffractive imaging},
  journal = {Ultramicroscopy},
  year    = {2009},
  volume  = {109},
  pages   = {1256--1262}
}

@article{miao2025,
  author  = {Miao, J.},
  title   = {Computational microscopy with coherent diffractive imaging and ptychography},
  journal = {Nature},
  year    = {2025},
  volume  = {637},
  pages   = {281--295}
}

@article{miao2016,
  author  = {Miao, J. and Ercius, P. and Billinge, S. J. L.},
  title   = {Atomic electron tomography: {3D} structures without crystals},
  journal = {Science},
  year    = {2016},
  volume  = {353},
  pages   = {aaf2157}
}

@article{muller2014,
  author  = {M{\"u}ller, K. and Krause, F. F. and B{\'e}ch{\'e}, A. and Schowalter, M. and Galioit, V. and L{\"o}ffler, S. and Verbeeck, J. and Zweck, J. and Schattschneider, P. and Rosenauer, A.},
  title   = {Atomic electric fields revealed by a quantum mechanical approach to electron picodiffraction},
  journal = {Nature Communications},
  year    = {2014},
  volume  = {5},
  pages   = {5653}
}

@article{nguyen2014,
  author  = {Nguyen, D. T. and Findlay, S. D. and Etheridge, J.},
  title   = {The spatial coherence function in scanning transmission electron microscopy and spectroscopy},
  journal = {Ultramicroscopy},
  year    = {2014},
  volume  = {146},
  pages   = {6--16}
}

@article{nogami2026,
  author  = {Nogami, M. and Seki, T. and Ooe, K. and Shibata, N.},
  title   = {Enhancing dose efficiency of optimum bright-field scanning transmission electron microscopy using a phase-shifted electron probe},
  journal = {Small Methods},
  year    = {2026},
  volume  = {10},
  pages   = {e01661}
}

@article{oleary2021,
  author  = {O'Leary, C. M. and Martinez, G. T. and Liberti, E. and Humphry, M. J. and Kirkland, A. I. and Nellist, P. D.},
  title   = {Contrast transfer and noise considerations in focused-probe electron ptychography},
  journal = {Ultramicroscopy},
  year    = {2021},
  volume  = {221},
  pages   = {113189}
}

@article{oleary2024,
  author  = {O'Leary, C. M. and Sha, H. and Zhang, J. and Su, C. and Kahn, S. and Jiang, H. and Zettl, A. and Ciston, J. and Miao, J.},
  title   = {Three-dimensional structure of buried heterointerfaces revealed by multislice ptychography},
  journal = {Physical Review Applied},
  year    = {2024},
  volume  = {22},
  pages   = {014016}
}

@article{ooe2019,
  author  = {Ooe, K. and Seki, T. and Ikuhara, Y. and Shibata, N.},
  title   = {High contrast {STEM} imaging for light elements by an annular segmented detector},
  journal = {Ultramicroscopy},
  year    = {2019},
  volume  = {202},
  pages   = {148--155}
}

@article{ooe2021,
  author  = {Ooe, K. and Seki, T. and Ikuhara, Y. and Shibata, N.},
  title   = {Ultra-high contrast {STEM} imaging for segmented/pixelated detectors by maximizing the signal-to-noise ratio},
  journal = {Ultramicroscopy},
  year    = {2021},
  volume  = {220},
  pages   = {113133}
}

@article{ooe2025,
  author  = {Ooe, K. and Seki, T. and Nogami, M. and Ikuhara, Y. and Shibata, N.},
  title   = {Dose-efficient phase-contrast imaging of thick weak phase objects via {OBF STEM} using a pixelated detector},
  journal = {Microscopy},
  year    = {2025},
  volume  = {74},
  pages   = {98--106}
}

@article{ooe2023,
  author  = {Ooe, K. and Seki, T. and Yoshida, K. and Kohno, Y. and Ikuhara, Y. and Shibata, N.},
  title   = {Direct imaging of local atomic structures in zeolite using optimum bright-field scanning transmission electron microscopy},
  journal = {Science Advances},
  year    = {2023},
  volume  = {9},
  pages   = {eadf6865}
}

@article{ophus2019,
  author  = {Ophus, C.},
  title   = {Four-dimensional scanning transmission electron microscopy ({4D-STEM}): from scanning nanodiffraction to ptychography and beyond},
  journal = {Microscopy and Microanalysis},
  year    = {2019},
  volume  = {25},
  pages   = {563--582}
}

@article{ophus2022,
  author  = {Ophus, C. and Zeltmann, S. E. and Bruefach, A. and Rakowski, A. and Savitzky, B. H. and Minor, A. M. and Scott, M. C.},
  title   = {Automated crystal orientation mapping in {py4DSTEM} using sparse correlation matching},
  journal = {Microscopy and Microanalysis},
  year    = {2022},
  volume  = {28},
  pages   = {390--403}
}

@article{oxley2020,
  author  = {Oxley, M. P. and Dyck, O. E.},
  title   = {The importance of temporal and spatial incoherence in quantitative interpretation of {4D-STEM}},
  journal = {Ultramicroscopy},
  year    = {2020},
  volume  = {215},
  pages   = {113015}
}

@article{ozdol2015,
  author  = {Ozdol, V. B. and Gammer, C. and Jin, X. G. and Ercius, P. and Ophus, C. and Ciston, J. and Minor, A. M.},
  title   = {Strain mapping at nanometer resolution using advanced nano-beam electron diffraction},
  journal = {Applied Physics Letters},
  year    = {2015},
  volume  = {106},
  pages   = {253107}
}

@article{pekin2017,
  author  = {Pekin, T. C. and Gammer, C. and Ciston, J. and Minor, A. M. and Ophus, C.},
  title   = {Optimizing disk registration algorithms for nanobeam electron diffraction strain mapping},
  journal = {Ultramicroscopy},
  year    = {2017},
  volume  = {176},
  pages   = {170--176}
}

@article{pennycook2015,
  author  = {Pennycook, T. J. and Lupini, A. R. and Yang, H. and Murfitt, M. F. and Jones, L. and Nellist, P. D.},
  title   = {Efficient phase contrast imaging in {STEM} using a pixelated detector. {Part 1}: Experimental demonstration at atomic resolution},
  journal = {Ultramicroscopy},
  year    = {2015},
  volume  = {151},
  pages   = {160--167}
}

@article{pennycook2019,
  author  = {Pennycook, T. J. and Martinez, G. T. and Nellist, P. D. and Meyer, J. C.},
  title   = {High dose efficiency atomic resolution imaging via electron ptychography},
  journal = {Ultramicroscopy},
  year    = {2019},
  volume  = {196},
  pages   = {131--135}
}

@book{press2007,
  author    = {Press, W. H. and Teukolsky, S. A. and Vetterling, W. T. and Flannery, B. P.},
  title     = {Numerical Recipes: The Art of Scientific Computing},
  edition   = {3rd},
  publisher = {Cambridge University Press},
  year      = {2007}
}

@incollection{rodenburg2008,
  author    = {Rodenburg, J. M.},
  title     = {Ptychography and related diffractive imaging methods},
  booktitle = {Advances in Imaging and Electron Physics},
  volume    = {150},
  pages     = {87--184},
  year      = {2008}
}

@article{schloz2025,
  author  = {Schloz, M. and Pekin, T. C. and Brown, H. G. and Byrne, D. O. and Esser, B. D. and Terzoudis-Lumsden, E. and Taniguchi, T. and Watanabe, K. and Findlay, S. D. and Haas, B. and Ciston, J. and Koch, C. T.},
  title   = {Improved three-dimensional reconstructions in electron ptychography through defocus series measurements},
  journal = {Microscopy and Microanalysis},
  year    = {2025},
  volume  = {31},
  pages   = {ozae110}
}

@article{seki2022,
  author  = {Seki, T. and Khare, K. and Murakami, Y. O. and Toyama, S. and S{\'a}nchez-Santolino, G. and Sasaki, H. and Findlay, S. D. and Petersen, T. C. and Ikuhara, Y. and Shibata, N.},
  title   = {Linear imaging theory for differential phase contrast and other phase imaging modes in scanning transmission electron microscopy},
  journal = {Ultramicroscopy},
  year    = {2022},
  volume  = {240},
  pages   = {113580}
}

@article{seki2018,
  author  = {Seki, T. and Takanashi, N. and Abe, E.},
  title   = {Integrated contrast-transfer-function for aberration-corrected phase-contrast {STEM}},
  journal = {Ultramicroscopy},
  year    = {2018},
  volume  = {194},
  pages   = {193--198}
}

@article{seki2018b,
  author  = {Seki, T. and Ikuhara, Y. and Shibata, N.},
  title   = {Theoretical framework of statistical noise in scanning transmission electron microscopy},
  journal = {Ultramicroscopy},
  year    = {2018},
  volume  = {193},
  pages   = {118--125}
}

@article{shen2025,
  author  = {Shen, Y. and Ooe, K. and Zhang, D. and Xie, L. and Zhang, H. and Kobayashi, S. and Seidel, J. and Shimakawa, Y. and Kan, D.},
  title   = {Ferroelectric {ZrO2/Hf0.5Zr0.5O2} epitaxial {Moir{\'e}} heterostructures},
  journal = {Applied Physics Letters},
  year    = {2025},
  volume  = {127},
  pages   = {112901}
}

@article{varnavides2026,
  author  = {Varnavides, G. and de Kleijne, W. P. M. and Ribet, S. M.},
  title   = {The {ABCs} of phase retrieval: Connecting the acronyms of scanning transmission electron microscopy},
  journal = {MRS Bulletin},
  year    = {2026},
  volume  = {51},
  pages   = {897--912}
}

@article{verbeeck2012,
  author  = {Verbeeck, J. and B{\'e}ch{\'e}, A. and Van den Broek, W.},
  title   = {A holographic method to measure the source size broadening in {STEM}},
  journal = {Ultramicroscopy},
  year    = {2012},
  volume  = {120},
  pages   = {35--40}
}

@article{xin2009,
  author  = {Xin, H. L. and Muller, D. A.},
  title   = {Aberration-corrected {ADF-STEM} depth sectioning and prospects for reliable {3D} imaging in {S/TEM}},
  journal = {Journal of Electron Microscopy},
  year    = {2009},
  volume  = {58},
  pages   = {157--165}
}

@article{yang2017,
  author  = {Yang, H. and MacLaren, I. and Jones, L. and Martinez, G. T. and Simson, M. and Huth, M. and Ryll, H. and Soltau, H. and Sagawa, R. and Kondo, Y. and Ophus, C. and Ercius, P. and Jin, L. and Kov{\'a}cs, A. and Nellist, P. D.},
  title   = {Electron ptychographic phase imaging of light elements in crystalline materials using {Wigner} distribution deconvolution},
  journal = {Ultramicroscopy},
  year    = {2017},
  volume  = {180},
  pages   = {173--179}
}

@article{yang2015,
  author  = {Yang, H. and Pennycook, T. J. and Nellist, P. D.},
  title   = {Efficient phase contrast imaging in {STEM} using a pixelated detector. {Part {II}}: Optimisation of imaging conditions},
  journal = {Ultramicroscopy},
  year    = {2015},
  volume  = {151},
  pages   = {232--239}
}

@article{yang2016,
  author  = {Yang, H. and Rutte, R. N. and Jones, L. and Simson, M. and Sagawa, R. and Ryll, H. and Huth, M. and Pennycook, T. J. and Green, M. L. H. and Soltau, H. and Kondo, Y. and Davis, B. G. and Nellist, P. D.},
  title   = {Simultaneous atomic-resolution electron ptychography and {Z}-contrast imaging of light and heavy elements in complex nanostructures},
  journal = {Nature Communications},
  year    = {2016},
  volume  = {7},
  pages   = {12532}
}

@article{zhou2019,
  author  = {Zhou, J. and Yang, Yongsoo and Yang, Yao and Kim, D. S. and Yuan, A. and Tian, X. and Ophus, C. and Sun, F. and Schmid, A. K. and Nathanson, M. and Heinz, H. and An, Q. and Zeng, H. and Ercius, P. and Miao, J.},
  title   = {Observing crystal nucleation in four dimensions using atomic electron tomography},
  journal = {Nature},
  year    = {2019},
  volume  = {570},
  pages   = {500--503}
}

\end{document}